\documentclass[prl,aps,twocolumn,floats]{revtex4}
\def\b{\begin{equation}}
\def\e{\end{equation}}

\begin{document}

\title
{Comment on ``Derivation of the Raychaudhuri Equation'' by Dadhich}



\author{Abhas Mitra}
 
\email {amitra@apsara.barc.ernet.in}
\altaffiliation{}
\affiliation
{Theoretical Astrophysics Section, Bhabha Atomic Research Center,
Mumbai-400085, India\\}


\begin{abstract}
In a recent preprint, gr-qc/0511123, Dadhich has given a brief yet
beautiful exposition on some of the research works by  Prof.
A.K. Raychaudhuri. Here Dadhich highlights the fact that the
apparently ``self-evident'' assumption of occurrence of ``trapped surfaces''
may not be realized  atleast in some  specific cosmological
models though no general proof  for non-occurrence of trapped surfaces
exists in the cosmological context. However, Dadhich added, without sufficient
justification, that trapped surfaces should occur for collapse of isolated bodies.
We point out that  actually trapped surfaces do not occur even for  collapse
of spherically symmetric isolated bodies. Further unlike the cosmological case, for isolated
bodies, an exact proof for generic non-occurrence of trapped surfaces is available.   Thus for
isolated bodies, the above referred apparently ``self-evident'' assumption fails
much more acutely than in cosmology. Many recent astrophysical observations tend
to corroborate the fact trapped surfaces do not occur for isolated bodies. Two
recent specific papers (PRD) are cited to show that when radiative non-diispative
collapse can prevent formation of trapped surfaces.
\end{abstract}

\pacs{Valid PACS appear here}
\maketitle

\section{Introduction} 
In a recent preprint entitled ``Derivation of the Raychaudhuri Equation''\cite{1},
Dadhich has presented a lucid and insightful rederivation of the celebrated
``Raychaudhuri Equation''. Dadich also briefly summarizes the recent works of Prof.
Raychoudhuri which considered specific examples/criteria of non-singular cosmologies. In recent times,
the first example of non-singular cosmologies came in 1990\cite{2} which involved
cylindrical geometry. As Dadhich emphasizes, these non-singular cosmological models
imply non-occurrence of trapped surfaces contrary to the crucial assumption behind singularity theorems.
It may be emphasized here, that, though there is no general proof for  non-occurrence of trapped surfaces in cosmological
context;  the singularity in the standard Friedmann universe may not necessarily
imply past trapped surface because it could be an artifact of assumed maximal
symmetry in the model.

 During this discussion, Dadhich, nonethess, writes that
 ``This assumption is quite justifiable for the case of collapse of an isolated body''.
It may be mentioned that the idea  of ``trapped surfaces''  appeared to be ``self-evident''
both for cosmology and isolated bodies till 1990 though in hindsight it may appear to be not so now
for cosmological context. In the following, we would show that though the idea of formation
of trapped surfaces appears to be ``justifiable'' for isolated bodies, actually, the situation here
is atleast as misleading as it was in cosmology prior to 1990.

This will be evident from the brief derivation presented below.

\section{The Proof}
Any spherically symmetric spacetime may be expressed as 
\begin{equation}
ds^2 = g_{00} dt^2 + g_{rr} dr^2 - R^2 (d\theta^2 + \sin^2\theta d\phi^2)
\end{equation}
where $\theta$ is polar, $\phi$ is azimuth  coordinate, and $R=R(r, t)$ is the  Circumference coordinate. 
  To start with, $r,t$ may be considered as arbitrary radial and time coordinates.
$R$ is also called ``areal'' or surface coordinate and is a scalar. Further, $R$  happens to be the physically observable {\em Luminosity Distance} (in a static universe). Any spherically
symmetric spacetime may be viewed as embedded with $R=fixed$ markers
against the background of which the fluid or the test particle moves.

For the specific case of the interior spacetime of a
 spherically symmertical fluid, we consider $r$ and $t$ as  the comoving
coordinates. For instance a marker $r=r_1$ signifies a certain mass shell
of the fluid containing fixed number of baryons and remains fixed by definition.
At a certain comoving time $t=t_1$, the surface area of this shell is $4\pi R_1^2(t)$
and $R_1$ is decreasing while $r_1$ stays fixed. This is the viewpoint for a comoving
observer something like that of the driver of a car who always finds the speed of the car to be
zero w.r.t. to him. However there could be milestones  and fixed speedometers on the road
who can find the car to be moving. Similarly, the fluid moves w.r.t. the background  grid of
$R=fixed$ markers.

For radial motion with $d\theta =d\phi =0$,
the metric becomes
\begin{equation}
ds^2 = g_{00} ~dt^2 (1- x^2) 
\end{equation}

where the auxiliary parameter 
\b
 x = {\sqrt {-g_{rr}} ~dr\over \sqrt{g_{00}}~ dt}
 \e
 Eq.(1) may be rewritten as
 \b
 (1-x^2) = {1\over g_{00}} {ds^2\over dt^2}
 \e
 Suppose an arbitrary roadside marker at a fixed $R$ is observing the fluid motion as the
 fluid passes by it.
    If we intend to find the parameter $x$ for such a $R=constant$
marker, i.e,  a roadside milestone at {\em fixed} $R$, we
will have,
\b
d R(r,t) = 0= {\dot R} dt + R^\prime dr 
\e
where an overdot denotes a partial derivative w.r.t. $t$ and a prime denotes
a partial derivative w.r.t. $r$.
Therefore, at a fixed $R$, we obtain,
\b
{dr\over dt} = - {{\dot R}\over R^\prime}
\e
and the corresponding $x$ is
\begin{equation}
x= x_{c} = {\sqrt {-g_{rr}} ~dr\over \sqrt{g_{00}}~ dt} = -{\sqrt {-g_{rr}}
~{\dot R}\over \sqrt{g_{00}}~ R^\prime}
 \end{equation}
Using Eqs.(3), we also have, 
\begin{equation}
(1-x_c^2) = {1\over g_{00}} {ds^2\over dt^2}
\end{equation}
Now let us define[3]
\begin{equation}
\Gamma = {R^\prime\over \sqrt {-g_{rr}}}
\end{equation}
\b
\qquad U = {{\dot R}\over \sqrt{g_{00}}}
\e
so that Eqs. (3) and (5)  yield
\b
x_c = {-U\over \Gamma}; \qquad U= -x_c \Gamma
\e
As is well known, the gravitational mass of the collapsing (or expanding) fluid is defined
through the equation[3]
\begin{equation}
\Gamma^2 = 1 + U^2 - {2M(r,t)\over R}
\end{equation}
Using Eq.(4) in (12) and then  transposing, we obtain
\begin{equation}
\Gamma^2 (1- x_c^2) = 1- {2M(r,t)\over R}
\end{equation}
By using Eqs.(8) and (9) in the foregoing Eq., we have
\begin{equation}
{{R^\prime}^2\over {-g_{rr} g_{00}}} {ds^2\over dt^2} = 1 - {2M(r,t)\over R}
\end{equation}
Recall that the determinant of the metric tensor is always negative:
$g = R^4 \sin^2 \theta ~g_{00} ~g_{rr} \le 0$,
so that we must always have
\b
-g_{rr}~ g_{00} \ge 0
\e
Further for the metric signature chosen here $ds^2 \ge 0$ for all material
particles or photons. Then it follows that the LHS of  Eq. (14) is {\em always positive}.
So must then be the RHS of the same Eq. and which implies that
\b
{2M(r,t)\over R} \le 1
\e
Since the choice of the $R=fixed$ marker is arbitrary ($0 <R <R_i$, where $R_i$
is the initial radius), the above result is a general one.
This shows, in a  most general fashion, that trapped surfaces
are not formed in spherical collapse or expansion of isolated bodies.

\section{Implications in Brief}
  If trapped surfaces are not formed then there is no guarantee that
  collapse results in a singularity. However, if one would insist
  that massive objects must collapse indefinitely because of existence
  of Chandrasekhar mass or Oppenheimer- Volkoff mass, ($M_{OV}$) i.e., if one would
  envisage $R\to 0$, Eq.(16) would demand that the gravitational mass of the
  final singular state is $M=0$. Immediately, the question would arise, then
  what is the nature of those compact objects with masses $M > M_{OV}$ found in
  many X-ray binaries and Active Galactic Nuclei? Although this small note is
  meant to show only non-occurrence of trapped surfaces for isolated bodies (Eq.[16]),
  we will make few comments with regard to the question posed above.
  
   Both Chandrasekhar mass and O-V mass refer to {\em cold} degenerate compact objects
   at temperature $T \approx 0$. On the other hand, if the compact object is
   composed of {\em hot} matter with immense radiation pressure, then they
   could be of arbitrary high mass like the fictitious Supermassive Stars.
  
  Dadhich writes that ``From the study of stellar structure we know that a sufficiently massive
body could, as its nuclear fuel exhausts, ultimately undergo indefinite collapse and therefore reaching
the trapped surface limit.''

 The above statement {\em ignores} the fact that even if there would be no nuclear fuel,
 a self-gravitating fluid  generates fresh source of internal energy and pressure
 by combination of Virial Theorem and Global Energy Conservation. This is the reason
 stellar mass proto stars and supermassive primordial clouds can survive millions
 of years without support of any nuclear burning. It is because of this effect, in
 reality, there cannot be any gravitational collapse without dissipation and heat/radiation transport. However general relativists more often than not ignore
 such physical aspects and instead consider textbook {\em adiabatic} collapse.
 In such a case, $M(r,t)$ either increases or remains fixed (at the boundary)
 and one happily obtains ``trapped surfaces''. In fact Govender \& Dadhich found
 that in some models of String Theories, gravitational collapse {\em necessarily
 generates radiation}\cite{4}. In classical GR too, same is true provided we properly
 incorporate
 physics in the problem. For dissipative collapse, $M(r,t)$ would decrease  with $R$ for all $r$
 and Eq.(16) must be obeyed.

Recently, Goswami \& Joshi[5] used the already known idea 
that loss of mass energy should prevent formation of trapped surfaces:

``The collapsing star radiates away most of its matter as the process of gravitational collapse evolves, so as to avoid the formation of trapped surfaces
and spacetime singularity''

However, the treatment of Goswami \& Joshi[5] is physically inconsistent because
they do not consider any radiation transport or dissipation at all! They 
unphysically and artificially
simulate decrease of $M(r,t)$ by considering an {\em adiabatic} collapse with a negative pressure.

On the other hand, there are genuine examples of non-occurrence of trapped surfaces
in the context of continued dissipative collapse in which radiation pressure and
energy density could grow unhindered. In brief, Santos \& Herrea[6] first showed
that effect of radiation pressure can not only stop the collapse but might even
caused a ``bounce''. And now, Herrera, Prisco \& Barreto[7] have numerically shown
that collapse ($ U <0$) of massive stars might turn into a bounce ($U >0$) because of growth of  radiation pressure in realistic dissipative collapse. Before ``bounce'' ($ U >0$ would occur, one must have a transition state with $U=0$. From
Eq.(12) such a state corresponds to
\begin{equation}
\Gamma^2 = 1  - {2M(r,t)\over R}
\end{equation}
Since $\Gamma^2 \ge 0$, one finds $2M(r,t)/R \le 1$, which implies occurrence of Eq.(16). If the collapse is reversed now, surely, there would not be any trapped surface. So, there is an explicit example where
radiation pressure can prevent formation of trapped surface. Eq.(16) nevertheless
holds true irrespective of existence specific examples.
\section{Conclusion}
   Collapse of isolated bodies is necessarily dissipative and in order that
   the worldlines of the collapsing fluid remains  non-spacelike, atleast for non-charged objects, it is
   necessary that trapped surfaces are not formed. However, in principle, an apparent
   horizon, $R(r,t) = 2 M(r,t)$ might form as $R \to 0$. But if one would work with
   the unphysical assumption of radiationless adiabatic collapse one would obtain
   trapped surfaces at finite $M$ and $R$. The radiation mentioned here refers to emission of neutrinos
   and photons and not Gravitational Radiation (since we are considering spherically
symmetric   evolution). 

In the absense of trapped surfaces, there would not be any
finite mass (uncharged) BH.
     There is already observational evidence that the so-called BH Candidates
     found in many X-ray binaries have strong intrinsic magnetic fields in lieu
     of any Event Horizon\cite{8}. Very recently, there is evidence that the
     compact object in the most well studied quasar Q0957+561 has similar properties\cite{9}.

 \end{document}